\documentclass[12pt,float,amsfonts,amsmath]{article}
\def \ed{\end{document}}
\usepackage{latexsym,epsfig,graphpap,graphics,amsmath}

\newtheorem{theorem}{Theorem}

\newtheorem{corollary}[theorem]{Corollary}

{\rm}

\newtheorem{remark}[theorem]{Remark}

\newenvironment{proof}[1][Proof]{\textbf{#1.} }{\
\rule{0.5em}{0.5em}}
\usepackage{amssymb}
\usepackage{epsfig}
\usepackage{amssymb}
\usepackage{latexsym}

\def \mr{\mathrm}

\def \b{\beta}

\def \inf{\infty}
\def \var{\varphi}

\def \z{\zeta}
\def \g1{\fr{1}{\Gamma(\al)}}

\def \sig{\sigma}
\def \Sig{\Sigma}

\def \bp{\begin{picture}}

\def \bth{\begin{theorem}}
\def \brem{\begin{remark}}
\def \erem{\end{remark}}
\def \bprf{\begin{proof}}
\def \eprf{\end{proof}}
\def \eth{\end{theorem}}
\def \bcor{\begin{corollary}}
\def \ecor{\end{corollary}}
\def \non{\nonumber\\}
\def \no1{\noindent}
\def \beq{\begin{eqnarray}}

\def \eeq{\end{eqnarray}}
\def \bi{\begin{itemize}}
\def \ei{\end{itemize}}

\def \ba{\begin{array}}

\def \ea{\end{array}}
\def \bt{\begin{tabular}}
\def \et{\end{tabular}}
\def \bc{\begin{center}}
\def \ec{\end{center}}
\def \mb{\mbox}
\def \un1{\underline}
\def \fr{\frac}
\def \l{\left}
\def \r{\right}
\def \al{\alpha}
\def \th{\theta}
\def \s1{\sqrt}

\def \Gam{\Gamma}
\def \hs1{\hspace*{5mm}}
\def \be{\begin{equation}}
\def \ee{\end{equation}}
\newcommand{\ds}{\displaystyle}
\def \vs1{\vspace{2mm}}
\def \v2{\vspace{7mm}}
\def \n1{\newpage}
\def \ra{\rightarrow}

\def \vs1{\vspace{2mm}}
\def \1{\begin{eqnarray}}
\def \2{\end{eqnarray}}
\def \3{\begin{eqnarray*}}
\def \4{\end{eqnarray*}}

\def \bn{\begin{enumerate}}
\def \en{\end{enumerate}}
\newcommand{\bqn}{\begin{equation}}
\newcommand{\9}{\end{equation}}
\begin{document}
\begin{titlepage}
\thispagestyle{empty}
\bigskip

\begin{center}
\noindent{\Large \textbf{ A Representation for the Anyon Integral
Function }} \\\vspace{1.5cm} {M. Aslam
Chaudhry\footnote{maslam@kfupm.edu.sa}, Amer
Iqbal\footnote{iqbal@math.washingtone.edu},
Asghar Qadir\footnote{ashqadir@kfupm.edu.sa, aqadirs@comsats.net.pk}}\\
\vspace{1cm}

${}^{1,3}${\it Department of Mathematical Sciences \\
King Fahd University of Petroleum and Minerals\\
Dhahran 31261, Saudi Arabia}

\vspace{3mm} ${}^2$
{\it School for Mathematical Sciences\\
Government College University\\
Lahore, 54600, Pakistan}

\vspace{3mm}${}^2$
{\it Department of Mathematics\\
University of Washington\\
Seattle, WA, 98195, U.S.A.}

\vspace{3mm}
${}^3${\it Centre for Advanced Mathematics and Physics\\
National University of Sciences and Technology\\
Campus of College of Electrical \& Mechanical Engineering
Rawalpindi, Pakistan}

\end{center}
\begin{abstract}
\baselineskip=18pt The Fermi-Dirac and Bose-Einstein particles
satisfy corresponding statistical distributions. In the phenomena of
charge fractionalization and the fractional quantum Hall effect it
is found that particles behave as if they are neither fermions nor
bosons. Such particles are called anyons. The integral functions for
bosons and fermions are available in the literature. However, there
is no anyon integral function available. In this note we propose a
pair of functions that interpolate, in some sense, between the gamma
function and the zeta function, which we call ``gamma-zeta"
functions. It is pointed out that this pair of functions very
naturally provides a representation of the anyon integral function.
\end{abstract}
\vspace{2cm}\begin{center}
\end{center}
\end{titlepage}
\newpage

\section{Introduction}
There are two aspects to this note: (1) the presentation of a new
special function; and (2) its physical application. It is
necessary to briefly review the background for both aspects, so
that readers familiar with one area can follow the motivation and
reasoning for the other. In this section, we will first recall the
special function aspect and then go on to explain the physical
aspect.

Most of the so-called ``special functions" arise as solutions of
commonly occurring first or second order linear differential
equations. However, two of the most significant special functions do
not come from differential equations, namely the gamma function and
the family of zeta functions. In fact both arose from the study of
numbers. (For a discussion of the relevant special functions, see
for example \cite{CZ}.) Considering their relevance and utility, and
the fact that they are related in the above sense, it seemed worth
while to try to put them together so that they appear as limiting
cases of either one of a pair of functions, which we call the {\it
gamma-zeta function} \cite{CQ}. It turns out that in doing so we
find that we can ``interpolate" between the Fermi-Dirac and
Bose-Einstein integral functions in a way that follows the physical
needs for doing so, as will be explained.

The gamma function was defined by Euler to extend the domain of the
factorial function from the integers and then the complex numbers.
It has the integral representation \bqn \Gamma(\alpha):= \int^\inf_0
e^{-t}t^{\alpha - 1}dt, \9 which is singular for negative integer
values of $\alpha$, but is well defined everywhere else. The zeta
function, defined over the reals, was first studied by Euler who
also gave a product formula for it known as the Euler product, \bqn
\zeta(s)=\sum_{n=1}^{\infty}\frac{1}{n^{s}}=\prod_{p:prime}(1-p^{-s})^{-1}\,.
\9 Riemann while studying the distribution of prime numbers
rediscovered the zeta function and studied it as a function defined
over complex numbers. He also established the amazing link between
the complex zeroes of zeta function and the distribution of prime
numbers. The Riemann zeta function, as it is now known, has the
integral representation \bqn \label{rep1}\z(\al): = \fr{1}{C (\al)}
\int^\inf_0 \fr{t^{\al-1}dt}{e^t + 1} \qquad (\al = \sig + i\tau,
\sig > 0), \9 where \bqn C(\al): = \Gam(\al) (1-2^{1-\al}).\9 The
function has the analytic continuation \cite{Dingle}\1
\label{rep2} \z(\al):= && \fr{1}{\Gam(\al) (e^{2\pi i\al}-1)}I(\al) \\
\nonumber && \fr{e^{-i\pi\al}\Gam(1-\al)}{2\pi i} I(\al), \2 where
\bqn I(\al): = \int_C \fr{z^{\al-1}dz}{e^z-1} \9 is the loop
integral. The contour $C$ consists of the real axis from $\inf$ to
$\rho$ $(0 < \rho < 2\pi)$, the circle $|z| = \rho$, and the axis
from $\rho$ to $\inf$. The integral $I(\al)$  is uniformly
convergent in any finite region of the complex plane and so defines
an integral function of $\al$. The representations \eqref{rep1} and
\eqref{rep2} provide an analytic continuation of $\z(\al)$ over the
whole complex plane. Since $I(1) = 2\pi i$ and $I(n) = 0 \quad
(n=2,3,\ldots)$, the poles of the gamma function $\Gam(1-\al)$ at
$\al = 2,3,4, \ldots$ cancel the zeros of $I(\al)$. It follows from
\eqref{rep2} that $\z(\al)$ has only a simple pole, due to
$\Gam(1-\al)$, at $\al=1$ with residue 1.

\section{Bose-Einstein and Fermi-Dirac Particles}
\setcounter{equation}{0} In quantum mechanics particles (for example
see \cite{QM}) are represented by wave functions that satisfy the
Schrodinger equation. Making the theory relativistic requires that
one shift over to quantum field theory (for example see \cite{QFT}).
Following the Dirac quantization procedure leads to the Klein-Gordon
equation, which has problems of interpretation due to the fact that
it has a second derivative relative to the time appearing in it. To
avoid these problems Dirac ``took the square root of the
Klein-Gordon equation" to obtain a first order equation in space and
time variables. It turned out that the Klein-Gordon equation
represents spin-less particles while the Dirac equation applies to
particles with spin. The theory further requires that the spin go up
in half-integer multiples of the quantity, $\hbar$, defined to be
$h/2{\pi}$, where $h$ is Planck's constant. The wave function for
identical particles with integer spin is symmetric while for
half-integer spin it is anti-symmetric under the exchange of
particles. On account of this property, particles with half-integer
spin can not have the same quantum numbers, while those with integer
spin can. The former are said to be subject to the {\it Pauli
exclusion principle}.

Statistical mechanics (for example see \cite{SP}) had been developed
to deal with ensembles of classical particles using the Maxwell
distribution: \bqn f(\varepsilon) = e^{(\mu - \varepsilon)/\tau}, \9
where $\tau$ stands for thermal energy and is hence related to the
temperature, $\varepsilon$ stands for the kinetic energy of the
given particle and $\mu$ for the chemical potential. The $f$ gives
the probability of the particle having the given kinetic energy. It
turned out that this distribution did not apply precisely to the
quantum particles but could be used as an approximation for a
mixture of the two types. Also, it was a good approximation for the
ensemble at high temperatures. For lower temperatures of systems of
particles of one type other distributions were required. For the
integer spin particles it was propounded by Bose and is called the
Bose-Einstein distribution: \bqn f_B(\varepsilon) =
\frac{1}{e^{(\varepsilon - \mu)/\tau} - 1}; \9 while for the
particles with half-integer spin it was propounded by Fermi and is
called the Fermi-Dirac distribution and is given by: \bqn
f_F(\varepsilon) = \frac{1}{e^{(\varepsilon - \mu)/\tau} + 1}. \9
Particles of the former type are called {\it bosons} and of the
latter type {\it fermions}.

The cumulative probabilities for bosons and fermions are given by
the Bose-Einstein and the Fermi-Dirac {\it integral
functions}:\bqn B_q(x): = \fr{1}{\Gam(q+1)} \int^\inf_0
\fr{t^q}{e^{ t- x} - 1} dt \qquad (q
> 0); \9 \bqn F_q(x): = \fr{1}{\Gam(q+1)} \int^\inf_0
\fr{t^q}{e^{t-x}+1}dt \qquad (q > -1). \9

\section{Anyons}
\setcounter{equation}{0}It appeared that all elementary particles
would belong to one of these classes. However, more recently, in
the phenomenon of the fractional quantum Hall effect \cite{ASW} it
was found that under certain conditions electron, a fermion, can
behave as if it is made of yet more fundamental particles with
spin a fractional multiple of $\frac{1}{2}\hbar$. Such particles
were dubbed "anyons" \cite{KR} (For a review see \cite{Wilczek}).
Particles such as anyons with spin/statistics interpolating
between bosons and fermions can only exits in two dimensions.
There are both topological and group theoretic reasons for their
non-existence in higher dimensions. As mentioned in the last
section, the wave function of bosons is symmetric under the
exchange of two particles whereas the wave function of the
fermions is antisymmetric. This has to do with the fact that in
three and higher dimensions the symmetry group is the permutation
group and it has only two one dimensional representations. The
trivial representation corresponds to the bosons and the
non-trivial representation gives fermions. In two dimensions,
however, the situation is much more interesting. The symmetry
group is larger than the permutation group, it is the braid
group\footnote{Braid group, $B_{N}$, is generated by
transpositions $\{\phi_{1},\cdots,\phi_{N}\}$ such that
$\phi_{i}\phi_{j}=\phi_{j}\phi_{i}$ for $|i-j|>1$ and
$\phi_{i}\phi_{i+1}\phi_{i}=\phi_{i+1}\phi_{i}\phi_{i+1}$. If we
further impose the relation $\phi_{i}^{2}=1$ then we get the
permutation group. It is the absence of this last relation in the
braid group which allows the possibility of anyons}. The braid
group has a one dimensional representation for every real number
$\nu$. Thus a two particle wave function under exchange of two
particles behaves in the following way:
\begin{eqnarray} \psi(b,a)=\left\{
                                       \begin{array}{ll}
                                         (+1)\,\psi(a,b), & \hbox{Bosons,} \\
                                         (-1)\,\psi(a,b), & \hbox{Fermions,} \\
                                         (-1)^{\nu}\psi(a,b), & \hbox{Anyons.}
                                       \end{array}
                                     \right.
\end{eqnarray}
There is no functional representation for anyons available in the
literature corresponding to that for bosons or fermions. The
gamma-zeta functions connect across the two in that for real $\nu$
they give an interpolation between them. As such, they should be
considered as candidates for representing the anyon integral
function. Note that $\Phi$ and $\Psi$ are sufficiently directly
related that either (or a linear combination of the two) could be
used. The choice would be guided by examining how convenient the
function is in the resulting formulae.

It was found \cite{TSG} that there are phenomena in which it
appeared that there were fractional charges appearing. This was
regarded as exciting because it seemed that these may be the
``quarks" that had been proposed in high energy physics
\cite{Gell-Mann}. However, the theory did not allow free quarks
\cite{GW} and it was found that the fractions did not necessarily
correspond to those required for quarks. This phenomenon was later
explained as arising from the {\it statistics} not being well
defined, namely that the particles behave like neither fermions nor
bosons but something in between. The explanation proposed can be
understood in terms of an example. Consider fermions of spin 1/2 (in
units of the Planck action $\hbar$) in an infinite linear array. The
axial component of the spin can then either be +1/2 or -1/2. The
energy is minimized by having them alternate in the axial component
of the spin. Had they been bosons there would be invariance under
shift by a lattice length. For fermions we would have to reverse the
sign on the shift. This can be achieved by shifting and then
``rotating the particle" through an angle $\pi$, i.e. changing the
phase of the particle wave function. This way we can have the
fermion behave like a boson in some sense. This is relevant for our
purposes. More generally the transformation may be through some
other fraction of $2\pi$.

We should be able to follow the same procedure for the integral
functions, of shifting the phase by $\pi$, or more generally, by
some other fraction of $2\pi$. Our proposed new ``gamma-zeta"
function has the Bose-Einstein and Fermi-Dirac integral functions
as limits. As such, it should be able to achieve just this. As
will be seen, the procedure adopted corresponds to the physical
argument just mentioned.

\section{The Gamma-Zeta Functions}
\setcounter{equation}{0} Our pair of new functions, which we call
{\it gamma-zeta functions}, is obtained by using the Mellin and Weyl
transforms. The Mellin transform of a function $\var(t)$, if it
exists, is defined by \cite{Zayed} \bqn \Phi_M(\al): = M[\var; \al]
= \int^\inf_0 t^{\al-1} \var(t)dt \qquad (\al = \sig + i\th).
\end{equation}
It is inverted by \bqn \var(t) = \fr{1}{2\pi i}
\int^{c+i\inf}_{c-i\inf} \Phi_M(z) t^{-z} dz.
\end{equation}
A convolution-like product of two functions $\var$ and $\psi$, $\var
\circ \psi$, can be defined by \cite{Debnath} \bqn (\var \circ
\psi)(t): = \int^\inf_0 \var(xt)\psi(x)dx.
\end{equation}
Note that the operation ``$\circ$'' is not commutative and
\bqn M[\var\circ \psi;\al] =
\Phi_M(\al)\Psi_M(1-\al),
\end{equation}
where
\bqn
\Phi_M(\al): = M[\var;\al]
\end{equation}
and
\bqn
\Psi_M(\al): = M[\psi;\al].
\end{equation}
It is to be noted that if
$\var\in L^1_{\mr loc} [0, \inf)$ is such that
\bqn
\var(t) = \l\{\ba{l}
\ds O(t^{-\sig_1}) \qquad (t\ra 0^+)\\
\ds O(t^{-\sig_2}) \qquad (t\ra \inf)
\ea \r.
\end{equation}
then $\Phi_M(\al)$ exists for $\sig_1 < \sig < \sig_2$. In
particular, if $\var$ is continuous on $[0, \inf)$ and has {\em
rapid decay} at infinity, the Mellin transform (4.1) will converge
absolutely for $\sig > 0.$

Let $X_M(\sig_1, \sig_2)$ be the space of all $\var\in L^1_{\mr
loc}[0, \inf)$ such that the corresponding integral (4.1) converges
uniformly and absolutely in the strip $\sig_1 < \sig < \sig_2$.
Then, for each $\var \in X_{M}(\sigma_{1}, \sigma_{2})$,
$\Phi_{M}(\alpha)$ is analytic in the interior of the strip
$\sigma_{1} \leq \sigma \leq \sigma_{2}$, \cite{Zayed}.

\vspace{4mm}

Similarly the Weyl transform of $\varphi \in X_{M} [0,\infty)$ is
given by \cite{Debnath}
\begin{equation}
\Phi_{W}(\alpha; x):=W^{-\alpha}[\varphi](x)=
\frac{1}{\Gamma(\alpha)} M[\varphi(x+t); \alpha] \qquad
(\alpha=\sigma+i\tau, \,\,\, \sigma > 0),
\end{equation}
where
\begin{equation}
\Phi_{W}(0, x) := \varphi(x),
\end{equation}
and
\begin{equation}
\Phi_{W}(-\alpha; x) := D^{m} (\Phi_{W}(\alpha; x)) \qquad
(\alpha=\sigma+i\tau, \,\,\, \sigma > 0),
\end{equation}
where $m$ is the smallest integer greater than $\sigma=Re(\alpha)$
and
\begin{equation}
D^{n} := (-1)^{n} \frac{d^{n}}{dx^{n}} \qquad (n=0,1,2, \ldots).
\end{equation}
In particular we have
\begin{equation}
\Phi_{W}(-n; x) = (-1)^{n} \varphi^{(n)} (x) \qquad (n = 0,1,2,
\ldots).
\end{equation}

For our purposes we need to consider some sufficiently well-behaved
functions. The class $\Sig (\sig_0)$ of ``good'' functions is
defined as follows: A function $\var\in \Sig(\sig_0)$ if \bn
\item[(P.1)] $\var\in X_M(0, \sig_0),$ \item[(P.2)] $\Phi_{W}(\al,x)$
is analytic in $x$ in the region $\mb{Re}(x) \geq 0$ for $0 \leq
\sig < \sig_0$. \en It is to be noted that the class $\Sig(\sig_0)$
is nonempty. To see this consider any polynomial $P(t)$. If $a > 0$,
the function $P(t) e^{-at}$ is a member of the class $\Sig(\sig_0)
\forall \sig_0 > 0$. Also, as $\Phi_{W}(0,x)$ is analytic in the
region $\mb{Re}(x) \geq 0$, each of $\var\in \Sig(\sig_0)$ is a
$C^\inf[0, \inf)$ function as well.

Let us now define \bqn \var_\nu(t):= \fr{e^{-\nu t}}{e^t-1} \9 and
\bqn \psi_\nu(t):= \fr{e^{-\nu t}}{e^t+1}, \qquad (\nu \geq 0, t >
0), \9 and consider their Weyl transforms: \bqn \Phi_\nu(\al;x): =
W^{-\al}[\var_\nu](x) \9 and \bqn \Psi_\nu(\al;x): =
W^{-\al}[\psi_\nu](x). \9 We call the functions $\Phi_\nu(\al;x)$
and $\Psi_\nu(\al;x)$ the {\it gamma-zeta functions}.  Note that \1
\label{gammazeta1} \Phi_\nu(\al;x) & = & \fr{1}{\Gam(\al)}
\int^\inf_0 t^{\al-1} \var_\nu(t+x)dt \\\nonumber & = & \fr{e^{-\nu
x}}{\Gam(\al)} \int^\inf_0 \fr{t^{\al-1} e^{-\nu t}}{e^{t+x} - 1}
dt\\\nonumber & = & \fr{e^{-(\nu+1)x}}{\Gam(\al)} \int^\inf_0
\fr{t^{\al-1} e^{-\nu t }}{e^t-e^{-x}} dt. \2 Similarly, we have
\bqn \label{gammazeta2} \Psi_\nu(\al;x) =
\fr{e^{-(\nu+1)x}}{\Gam(\al)} \int^\inf_0 \fr{t^{\al-1} e^{-\nu
t}}{e^t + e^{-x}} dt. \9 From \eqref{gammazeta1} and
\eqref{gammazeta2} we find that the gamma-zeta function $\Phi_
\nu(\al;x)$ is related to the general Hurwitz-Lerch zeta function
\cite{EMOT} $\Phi(z,s,\nu)$ via \bqn \Phi_\nu(\al;x) =
e^{-(\nu+1)x}\Phi(e^{-x},\al,\nu+1) ,\9 which leads to the series
representation \1 \label{lerch1}&& \Phi_\nu(\al;x) = e^{-(\nu+1)x}
\sum^\inf_{n=0} \fr{e^{-nx}}{(n+\nu+1)^\al} \non && \qquad (\nu\neq
-1,-2,-3,\ldots; \mb{ Re}(x) > 0; x = 0, \sig > 1). \2 Similarly, we
have the relation \bqn \Psi_\nu(\al;x) = e^{-(\nu+1)x}
\Phi(-e^{-x},\al,\nu+1) \9 that leads to the series representation
\1 \label{lerch2}&& \Psi_\nu(\al;x) = e^{-(\nu+1)x} \sum^\inf_{n=0}
\fr{(-1)^n e^{-nx}}{(n+\nu+1)^\al} \non && \qquad (\nu\neq -1, -2,
-3, \ldots; \mb{ Re}(x) > 0; x = 0, \sig > 0). \2 \bth The functions
$\Phi_\nu(\al;x)$ and $\Psi_\nu(\al;x)$ are related via \bqn
\Psi_{\nu}(\al;x) = (-1)^{(\nu + 1)}\Phi_\nu(\al;x + i \pi). \9 \eth
\bprf Replace $x$ by $x + i\pi$ in \eqref{lerch1} and compare with
\eqref{lerch2} to obtain the result. \eprf

Another useful relation, also proved elsewhere \cite{CQ} is: \bth
The functions $\Phi_\nu(\al;x)$ and $\Psi_\nu(\al;x)$ are related
via \bqn \label{rel2} \Psi_{2\nu} (\al;x) = \Phi_{2\nu} (\al;x) -
2^{1-\al} \Phi_\nu(\al;2x). \9 \eth

{\bf Proof:} \begin{eqnarray}
e^{(2\nu+1)x}(\Psi_{2\nu}(\alpha:x)-\Phi_{2\nu}(\alpha;x))&=&\sum_{n\geq
0}\frac{e^{-nx}-(-1)^{n}e^{-nx}}{(n+1+2\nu)^{\alpha}}\\\nonumber
&=&\sum_{k\geq
0}\frac{2e^{-(2k+1)x}}{(2k+2+2\nu)^{\alpha}}\\\nonumber &=&
2^{1-\alpha}e^{-x}\sum_{k\geq
0}\frac{e^{-2kx}}{(n+1+\nu)^{\alpha}}\\\nonumber &=&
2^{1-\alpha}e^{-x}\Phi_{\nu}(\alpha;2x)e^{(\nu+1)2x}
\end{eqnarray}

Note that our gamma-zeta functions are {\it dual} to each other in
the sense that the above relation can be easily inverted, so that
each is similarly related to the other. This fact is of special
relevance for us.

\section{The Anyon Integral Functions}
\setcounter{equation}{0} From the integral representations
\eqref{gammazeta1} of $\Phi_\nu(\al;x)$ and (2.4) of the
Bose-Einstein integral function $B_\al(x)$, we find that \bqn
B_{\al-1} (-x) = \Phi_0(\al;x), \9 leading to the fact that the
gamma-zeta function $\Phi_\nu(\al;x)$ is a natural extension of the
Bose-Einstein integral function. Similarly, from \eqref{gammazeta2}
and (2.5), we find that \bqn F_{\al-1} (-x) = \Psi_0(\al;x), \9
which shows that the second gamma-zeta function $\Psi_\nu(\al;x)$
naturally extends the Fermi-Dirac integral function.

Putting $\nu = 0$ and replacing $x$ by $-x$ and $\al$ by $\al+1$ in
\eqref{rel2} we find that the Fermi-Dirac and Bose-Einstein
integrals are related via \bqn F_\al(x) = B_\al(x) - 2^{1-\al}
B_\al(2x). \9 The relation (5.3) does not seem to have been realized
earlier.

Using the fractional Weyl transform $W^{-\alpha}$ on
$\Phi_\nu(\beta; t)$ we obtain \bqn \Phi_\nu(\alpha + \beta; x) =
\fr{1}{\Gam(\al)} \int^\inf_0 t^{\al-1} \Phi_\nu(\beta; t + x). \9
Putting $\nu = 0$ above we find the interesting integral
representation \bqn B_{\al+\b-1} (x) = \fr{1}{\Gam(\al)}
\int^\inf_0 t^{\al-1} B_{\b-1} (x-t) dt \9 for the Bose-Einstein
integral. Similarly, we obtain \bqn F_{\al+\b-1} (x) =
\fr{1}{\Gam(\al)} \int^\inf_0 t^{\al-1} F_{\b-1} (x-t) dt. \9
Putting $\al=1$ in (5.5) and (5.6) we obtain \bqn B_\b(x) =
\int^\inf_0 B_{\b-1} (x-t)dt, \9 and \bqn F_\b(x) = \int^\inf_0
F_{\b-1} (x-t)dt. \9

These representations for the Bose-Einstein and Fermi-Dirac
integral functions may prove useful.

We can interpolate between the Bose-Einstein and Fermi-Dirac
integral functions using {\it gamma-zeta} function. To see this
consider $G_{\nu}(\alpha; x)$, \begin{equation}
G_{\nu}(\alpha,x)=a(\nu)\Phi_{\nu}(\alpha;x)+b(\nu)\Psi_{\nu-1}(\alpha;x)\,.\end{equation}
If we appropriately chose $a(\nu), b(\nu)$ satisfying the conditions
$a(0)=1-b(0)=1$ and $a(1)=b(1)-1=0$ then $G_{\nu}(\alpha;x)$
interpolates between the Bose-Einstein and Fermi-Dirac integral
functions.

\section{Concluding Remarks}
The family of zeta functions including the Riemann, Hurwitz,
Epstein, Lerch, Selberg and their generalizations, constantly find
new applications in different areas of mathematics (number theory,
analysis, numerical methods, etc.) and physics (quantum field
theory, string theory, cosmology, etc.) A useful generalization of
the family is expected to have wide applications in all these areas
as well.  On the mathematical side, the gamma-zeta functions
$\Phi_\nu(\al;x)$ and $\Psi_\nu(\al;x)$ discussed in this paper
provide a unified approach to the study of the zeta family that is
remarkably simple. This is discussed in more detail in a separate
paper \cite{CQ}. For our present purposes of the physical
applications, the first pair of gamma-zeta functions also provide a
representation of an {\it anyon} integral function. There was no
such representation available in the literature. They also led to a
new relation between the Bose-Einstein and Fermi-Dirac integral
functions that reflect the nature of the two functions. The
``mixing" of fermions and bosons displayed in this relationship
demonstrates how the fractional spin behaviour arises.

One might have hoped that the gamma-zeta functions could be applied
to supersymmetry \cite{WZ}, as they ``unify" bosons and fermions in
some sense. Despite the ``duality" manifested by them, they cannot
be so used. The reason is that supersymmetry does not {\it
interpolate} between the two types of particles. They remain
quantized with spin whole or half integer {\it with nothing in
between}. The utility of the gamma-zeta functions is that they {\it
do} provide such an interpolation. This lack of intermediate
particles is in sharp contradistinction to the situation for grand
unified theories \cite{GG}, in which quarks and leptons are unified.
In this case there {\it are} intermediate particles, called {\it
lepto-quarks} that are neither quarks nor leptons but both. Such
particles are required for spontaneous symmetry breaking of the
unified symmetry \cite{PS}. This highlights the problem with
supersymmetry, that it does not have a spontaneous symmetry breaking
mechanism and has to use ``soft symmetry breaking". Again, one sees
a strong relationship between the physical and mathematical
requirements in the context of our gamma-zeta functions.

It is expected that the operational properties of these functions
can and will be exploited further in solving difficult problems in
mathematics and physics.

\v2
 \noindent {\bf Acknowledgment}: Two of the authors are grateful
to the King Fahd University of Petroleum \& Minerals for the
research facilities through the Research Project MZ/Zeta/242.

\end{document}